\documentclass[11pt,a4paper]{article}
\usepackage{bm}
\usepackage{latexsym}
\usepackage{dcolumn}
\usepackage{amsfonts,amssymb}
\usepackage{calc}
\usepackage{ifthen}
\usepackage{enumerate}
\usepackage{graphicx}
\usepackage[dvips]{epsfig}
\usepackage{latexsym,amssymb}
\usepackage{amsmath}
\usepackage{amstext}
\def\beq{\begin{equation}}
\def\eeq{\end{equation}}
\def\bea{\begin{eqnarray}}
\def\eea{\end{eqnarray}}
\def\benu{\begin{enumerate}}
\def\eenu{\end{enumerate}}
\def\nn{\nonumber}

\def\pa{{\partial}}
\def\l{\left}
\def\r{\right}
\def\lp{L_{_{\rm P}}}

\def\d{{\rm d}}
\begin{document}

%%%%%%%%%%%%%%%%%%%%%%%%%%%%%%%%%%%%%%%%%%%%%%%%%%%%%%%%%%%%%%%%%%%%%%%%%%%%%%%

\thispagestyle{plain}

%%%%%%%%%%%%%%%%%%%%%%%%%%%%%%%%%%%%%%%%%%%%%%%%%%%%%%%%%%%%%%%%%%%%%%%%%%%%%%%
\begin{center}
\Large{\bf Path integral duality modified propagators
\vskip 2pt
in spacetimes with constant curvature}
\vskip 8pt
\large{Dawood Kothawala$^{1+}$, L.~Sriramkumar$^{2\dag}$,
\vskip 2pt
S.~Shankaranarayanan$^{3\ddag}$ and T.~Padmanabhan$^{1\ast}$}
\vskip 4pt
$^{1}${\small \it IUCAA, Post Bag 4, Ganeshkhind, Pune 411 007, 
India.}
\vskip 2pt
$^{2}${\small \it Harish-Chandra Research Institute, Chhatnag 
Road, Jhunsi, Allahabad~211~019, India.}
\vskip 4pt
$^{3}${\small \it Institute of Cosmology and Gravitation, 
University of Portsmouth,\\ 
Mercantile House, Portsmouth~P01 2EG, U.K..}
\vskip 4pt
\hspace{-15mm}{\small \tt E-mail}:~$^{+}${\small \tt dawood@iucaa.ernet.in}, 
$^{\dag}${\small \tt sriram@hri.res.in},\\
\hspace{18mm}$^{\ddag}${\small \tt Shanki.Subramaniam@port.ac.uk},
$^{\ast}${\small \tt paddy@iucaa.ernet.in}
\end{center}
%%%%%%%%%%%%%%%%%%%%%%%%%%%%%%%%%%%%%%%%%%%%%%%%%%%%%%%%%%%%%%%%%%%%%%%%%%%%%%%
\centerline{\small (\today)}
%%%%%%%%%%%%%%%%%%%%%%%%%%%%%%%%%%%%%%%%%%%%%%%%%%%%%%%%%%%%%%%%%%%%%%%%%%%%%%%
\begin{abstract}
The hypothesis of path integral duality provides a prescription to 
evaluate the propagator of a free, quantum scalar field in a given 
classical background, taking into account the existence of a 
fundamental length, say, the Planck length, $\lp$, in a {\it locally 
Lorentz invariant manner}.\/ 
We use this prescription to evaluate the duality modified propagators in spacetimes with {\it constant curvature}\/ (exactly in the case of
one spacetime, and in the Gaussian approximation for another two), and 
show that:~(i)~the modified propagators are ultra violet finite, (ii)~the 
modifications are {\it non-perturbative}\/ in $\lp$, and (iii)~$\lp$ seems 
to behave like a `zero point length' of spacetime intervals such that 
$\l\langle \sigma^2(x,x')\r\rangle = \l[\sigma^{2}(x,x')+ {\cal O}(1)\; 
\lp^2 \r]$, where $\sigma(x,x')$ is the geodesic distance between the two 
spacetime points $x$ and~$x'$, and the angular brackets denote (a suitable) 
average over the quantum gravitational fluctuations. 
We briefly discuss the implications of our results.
\end{abstract} 
\vskip 12pt
%%%%%%%%%%%%%%%%%%%%%%%%%%%%%%%%%%%%%%%%%%%%%%%%%%%%%%%%%%%%%%%%%%%%%%%%%%%%%%%
%{\small PACS:}{\small 04.62.+v,04.60.-m} 
\vskip 12pt
%%%%%%%%%%%%%%%%%%%%%%%%%%%%%%%%%%%%%%%%%%%%%%%%%%%%%%%%%%%%%%%%%%%%%%%%%%%%%%%

\section{Motivation}

It is presumed that quantum gravitational effects would become important 
at length scales of the order of the Planck length, $\lp = (G\, \hbar 
/c^3)^{1/2}$. 
At these scales, it seems quite likely that the description of the spacetime 
structure in terms of a metric, as well as certain notions of standard quantum 
field theory, would have to undergo drastic changes. 
Since any quantum field has virtual excitations of arbitrary high energy, 
which probe arbitrarily small scales, it follows that the conventional 
quantum field theory can only be an approximate description that is valid 
at energies smaller than the Planck energy.

The existence of a fundamental length implies that processes involving 
energies higher than the Planck energy can possibly be suppressed, 
thereby improving the ultra violet behavior of the theory. 
In particular, one hopes that, in a complete theory, gravity would provide 
an effective cut off at the Planck scales~\cite{dewitt-1964}. 
Some very general considerations based on the principle of equivalence and 
the uncertainty principle seem to strongly indicate that it may not be 
possible to operationally define spacetime events beyond an accuracy of the 
order of~$\lp$.
Therefore, one may consider $\lp$ as the `zero point length' of spacetime 
intervals~\cite{paddy-1985}. 
Specifically, if $\sigma(x,x'|g_{\mu\nu})$ denotes the geodesic distance 
between the spacetime points $x$ and $x'$ in the background metric 
$g_{\mu\nu}$, then one can expect that
\beq
\lim_{x \rightarrow x'}  
\l\langle\sigma^2(x,x'\vert g_{\mu\nu} + h_{\mu\nu})\r\rangle 
= \mathcal{O}(1)\; \lp^2,\label{eq:zpl-coinc}
\eeq
where $h_{\mu\nu}$ represents {\it all}\/ possible quantum fluctuations 
about the background metric, and the angular brackets represent a suitable 
path integral average over these fluctuations. 
Such a behavior can then be expected to render the coincidence limit of 
the propagators finite.

Let us now briefly outline as to how such a result might arise in a 
simple example. 
Consider a toy model of quantum gravity, where one considers {\it quantum 
fluctuations}\/ in the conformal factor of the metric of flat spacetime. 
In such a case
\beq
g_{\mu \nu}(x)=\l[ 1 + \varphi(x) \r]^2\; \eta_{\mu\nu},
\eeq
and one treats $\varphi(x)$ as the relevant degrees of freedom describing 
the spacetime, which must be quantized. 
(It is often claimed in literature that the conformal factor should 
not be regarded as a `physical' degree of freedom. This is incorrect, and we have outlined the essential arguments in 
App.~\ref{app:3d-conformal}.) 
It can be easily shown that \cite{jvn-paddy-book}
\beq
\l\langle \varphi(x)\; \varphi(x')\r\rangle 
= \frac{\lp^2}{\sigma_{_{\rm M}}^{2}\l(x,x'\r)},
\label{eq:qcf-corr}
\eeq
where $\sigma_{_{\rm M}}^{2}(x,x')$ is the Lorentz invariant interval in 
flat spacetime. 
We shall now explicitly illustrate how a result such as~(\ref{eq:zpl-coinc}) 
might be obtained. 
We write down the Taylor series expansion of the Synge's world function, 
$\Omega(x,x') = [\sigma^2(x,x')/2]$ around the base point, say, $x$, in 
terms of the parameter~$s$. 
If we now take the vacuum expectation value 
$\underset{x \rightarrow x'}{\mathrm{lim}} \l\langle \Omega (x, x')\r\rangle$, 
and use Eq.~(\ref{eq:qcf-corr}), each term of the expansion turns out to be 
proportional to $\lp^2$. 
(Specifically, after point-splitting the resulting series, the 
$k^{\rm th}$ term contains $(k-2)$ derivatives of the right hand 
side of Eq.~(\ref{eq:qcf-corr}), which is multiplied by $(x-x')^k$, 
thereby giving a finite coincidence limit.) 
In fact, an {\it exact}\/ path integral average of the propagator, say, $G(x,x')$, over the quantum fluctuations of $\varphi(x)$ can be carried out 
for the special case of a conformally coupled scalar field~\cite{paddy-1985}. 
The result obtained is
\beq
\langle G[\sigma(x,x')]\rangle 
= G\l(\langle \sigma(x,x'\rangle\r)
\eeq
with $\langle \sigma (x,x')\rangle$ interpreted as in Eq.~(\ref{eq:zpl-general}) 
below. 
These calculations suggest that Eqs.~(\ref{eq:zpl-coinc}) 
and~(\ref{eq:zpl-general}) might be valid in a more general context. 
As we shall see, the results we obtain in this paper indeed point to such a 
conclusion (as was originally suggested in Ref.~\cite{paddy-1997-1998}).

The points mentioned above are highly suggestive. 
However, to set them on a more firm basis, one must use a 
physically well motivated hypothesis that incorporates one 
or more of the effects expected to arise in the actual 
quantum theory of gravitation.
It would then indeed be interesting to examine whether one could 
derive a connection between the results mentioned above and some 
general underlying principle. 
Perhaps, the best strategy would be to accept our ignorance of 
the Planck scale physics and use, at the semiclassical level, a 
prescription to obtain effective quantum field theoretic propagators 
which captures the resultant smearing of quantum fields over a region 
of size of the order of~$\lp$. 
One such prescription was outlined in an earlier work, wherein 
the invariance of the path integral for a free, relativistic 
particle, under a {\it duality transformation}\/ which keeps $\lp$ 
invariant (for more details, see the next section) was 
demanded, and the corresponding modifications to the two point 
function was worked out~\cite{paddy-1997-1998}. 
It was shown that, in the Minkowski spacetime, the hypothesis of path 
integral duality (PID) proves to be essentially equivalent to adding a 
zero point length to spacetime intervals, in the sense of 
Eq.~(\ref{eq:zpl-coinc})---a relation, which we should stress is not at 
all obvious. 
The PID approach has since been utilized to compute the possible quantum
gravitational modifications to a variety of effects in flat as well as 
curved spacetimes~\cite{srini-1998,shanki-2001,sriram-2006,dawood-2008}.

In this paper, we shall consider the PID modifications to the two point 
function of a quantum scalar field in spacetimes with constant scalar 
curvature. 
After a rapid outline of the PID prescription, we shall evaluate the
modified Green's function in the Einstein static universe and the de Sitter 
as well as the anti-de Sitter spacetimes. 
We find that, in $(3+1)$-dimensions, while the modified propagator can be 
evaluated exactly in the Einstein static universe, it is not amenable to
an exact calculation in the de Sitter and the anti-de Sitter spacetimes. Therefore, we compute the modified propagator in the Gaussian approximation 
in the latter two cases.
In the case of the anti-de Sitter spacetime, it turns out that 
an exact evaluation of the modified propagator is possible in 
$(2+1)$-dimensions~\cite{dawood-2008}, an example which we shall 
also briefly consider.
We shall show that, in all these cases, the modified propagator 
remains finite in the coincidence limit, and the modifications are 
non-perturbative in $\lp$. 
The form of the modified propagators that we obtain suggests that, under the 
PID prescription, the geodesic distance $\sigma(x,x' \vert g_{\mu\nu})$ is 
modified as
\beq
\l\langle \sigma^2(x,x'\vert g_{\mu\nu} 
+ h_{\mu\nu})\r \rangle 
= \sigma^2(x,x'\vert g_{\mu\nu}) 
+ {\cal O}(1)\; \lp^2, \label{eq:zpl-general}
\eeq
in agreement with Eq.~(\ref{eq:zpl-coinc}).

We shall work with the metric signature $(-, +, +, \ldots, +)$, and 
we shall set $\hbar$ and $c$ to unity.
Also, we shall denote the set of coordinates $x^{\mu}$ as $x$.
Moreover, for convenience, we shall hereafter refer to the geodesic 
distance $\sigma(x, x' \vert g_{\mu\nu})$ simply as $\sigma(x,x')$.

%%%%%%%%%%%%%%%%%%%%%%%%%%%%%%%%%%%%%%%%%%%%%%%%%%%%%%%%%%%%%%%%%%%%%%%%%%%%%%%

\section{The setup}\label{sec:su}

Let $G(x, x')$ denote the two point function associated with a quantized, 
free scalar field that is propagating in a classical gravitational 
background described by the metric tensor~$g_{\mu\nu}$. 
The two point function can be expressed as a relativistic path integral 
with the action given by~$[m\, {\mathcal R}(x,x')]$, 
where $m$ is the mass of the scalar field and ${\mathcal R}(x,
x')$ is the proper length of the path linking the two spacetime 
points ${x}$ and ${x'}$ (see, for instance, 
Ref.~\cite{jvn-paddy-book}). 
The hypothesis of PID suggests that the path integral amplitude in 
the sum over the paths be modified such that paths with lengths below 
the Planck scale are suppressed in the sum, while maintaining invariance 
under the {\it duality transformation}:\/ ${\mathcal R} \rightarrow 
\l(\lp^2/{\mathcal R}\r)$. 
The specific prescription being that, instead of the original sum over 
paths, viz.
\beq
G(x,x')
= \underset{\rm{all~paths}}{\sum}
\exp\, i\, \l[m\; {\mathcal R}(x, x')\r],
\eeq
one considers the following~\cite{paddy-1997-1998}:
\beq
G_{_{\rm PID}}(x,x')
=\underset{\rm{all~paths}}{\sum} 
\exp\, {i\, m\,\l[{\mathcal R} (x, x')
+ \frac{\lp^2}{{\cal R}(x, x')}\r]}.
\eeq
In the Minkowski spacetime, such a modified sum over the paths can be 
carried out exactly using lattice techniques~\cite{paddy-1997-1998}.  
For a massless scalar field, it can be shown that the modified propagator
is given by
\beq 
G_{_{\rm PID}}(x,x') 
= \l(\frac{i}{4\,\pi^{2}}\r)\, 
\frac{1}{\sigma_{_{\rm M}}^{2}(x,x') + \lp^2},
\label{eq:mGfnmst}
\eeq 
where, as we mentioned earlier, $\sigma_{_{\rm M}}^{2}(x,x')$
denotes the Lorentz invariant, flat spacetime interval.
Evidently, the above PID modified propagator is finite in the coincidence 
limit. 
Also, the form is indeed suggestive of the spacetime itself as 
having a minimal length scale of ${\mathcal O}(\lp)$---a feature, 
as we had discussed, that is expected to arise when we take the 
quantum gravitational effects into account.
One must note that {\it it is not a priori evident as to why there 
must exist any connection whatsoever between the PID invariance and 
the existence of a minimal length scale}.\/ 
Hence, as it has been stressed in earlier works, the result in 
Eq.~(\ref{eq:mGfnmst}) is non-trivial. 
But, we should add that a similar result arrived at using the T-duality 
in string theory seems to indicate towards the possibility of such a 
connection~\cite{fontanini-2006}. 
Returning to the modified propagator~(\ref{eq:mGfnmst}), we see 
that it is {\it non-perturbative}\/ in~$\lp$. 
Clearly, one could not have obtained this result by a perturbation 
expansion in~$\lp$.

To apply the above results to curved spacetime, let us recall that 
the propagator $G(x, x')$ can also be expressed in 
Schwinger's proper time representation as 
follows~\cite{schwinger-1951,dewitt-1975}:
\beq
G(x,x') 
=i\, \int\limits_0^{\infty}\d s\; K(x, x'; s), 
\label{eq:Gfn} 
\eeq 
where $K(x,x'; s)$ is given by
\beq
K(x, x'; s) 
\equiv \langle x \vert \exp{- \l(i\, {\hat {\cal H}}\, s\r)} 
\vert x' \rangle.
\eeq 
The quantity $K(x, x'; s)$ is formally equivalent to 
the path integral amplitude for a quantum mechanical system described 
by the Hamiltonian
\beq
{\hat {\cal H}} 
= -\l(\Box - m^{2} - \xi\, R\r)
\equiv -\frac{1}{\sqrt{-g}}\,
\partial_{\mu}\l(\sqrt{-g}\, g^{\mu\nu}\, \partial_{\nu}\r)
+m^2+ \xi\, R 
\eeq
with $\xi$ being the coefficient of non-minimal coupling and $R$ denoting
the scalar curvature of the background spacetime.
Note that, in the kernel $K(x, x'; s)$, $s$ plays the 
role of the `time' parameter (the so called `proper time'). 
In flat spacetime, it was shown that PID modifies the 
expression~(\ref{eq:Gfn}) for the two point function to 
be~\cite{paddy-1997-1998}
\beq
G_{_{\rm PID}}(x,x') 
= i\,\int\limits_0^{\infty} \d s \;\, {\rm e}^{i\, \lp^2/4\,s}\; 
K(x, x'; s).
\label{eq:mGfn}
\eeq
As mentioned above, such a modification has also been derived in the 
context of string theory using T-duality~\cite{fontanini-2006}. 
While this connection is very suggestive, it is \textit{not}\/ a rigorous 
proof and, in this paper, we shall treat Eq.~(\ref{eq:mGfn}) as a 
prescription and explore the consequences. 

In $(D+1)$~spacetime dimensions, the proper time kernel $K(x, x'; s)$ 
can be expressed in the Schwinger-DeWitt series representation as follows \cite{dewitt-1975,parker-cargese,birrell-1982,camporesi-1990}:
\beq
K(x, x'; s) 
= i\; (4\,\pi\, i\, s)^{-(D+1)/2}\;\, 
\Delta^{1/2}(x, x')\;\, 
{\rm e}^{i\, \sigma^{2}(x,x')/4\,s}\;\, 
\sum \limits_{n=0}^{\infty}\; a_{n}(x, x')\;\, 
(i\,s)^{n},\label{eq:sde} 
\eeq
where  $\sigma^{2}(x,x')$ is the square of the proper distance along the geodesic between the spacetime points ${x}$ and ${x'}$.
The quantity $\Delta(x, x')$ is a biscalar that is
defined as
\beq
\Delta(x, x')
=-\, \l[-g(x)\r]^{-1/2}\;\, {\rm Det}\; 
\l(-\frac{\pa^2 \l[\sigma^{2}(x,x')/2\r]}{\pa x^{\mu}\,
\pa x'^{\nu}}\r)\;\, \l[-g(x')\r]^{-1/2}
\eeq
with $g(x)$ being the determinant of the metric tensor $g_{\mu\nu}$.
It can be shown that the coefficients $a_{n}$ satisfy the recursion 
relation \cite{camporesi-1990}
\beq 
a_{n}(x, x') 
= \sigma^{-n}(x, x')\;
\int\limits_{0}^{\sigma}\; \d\sigma(q)\; \sigma^{(n-1)}(q)\;\,
\l[\Delta^{-1/2}\; \square\, \l(\Delta^{1/2}\; a_{n-1}\r)\r][x(q),x'],
\label{eq:rr} 
\eeq
where $\sigma(q) = \sigma\l[x(q),x'\r],$ and the integral is along the geodesic connecting the spacetime points $x$ and $x'$.
In what follows, we shall use the above representation of the kernel
in Eq.~(\ref{eq:mGfn}) to evaluate the PID modified propagator in 
spacetimes of constant curvature.

%%%%%%%%%%%%%%%%%%%%%%%%%%%%%%%%%%%%%%%%%%%%%%%%%%%%%%%%%%%%%%%%%%%%%%%%%%%%%%%%

\section{The modified propagator in spacetimes with constant curvature}

Spacetimes with constant scalar curvature $R$ are maximally symmetric.
Let $\ell$ be a constant that denotes the characteristic length scale associated with these spacetimes.
Then, the metric describing these spacetimes can be expressed as (see,
for instance, Ref.~\cite{weinberg-1972})
\beq
\d s^2 
= g_{\mu\nu}\,  \d x^{\mu}\, \d x^{\nu}
= \eta_{\mu\nu}\, \d x^{\mu}\, \d x^{\nu} 
+ \l(\frac{{\cal K}/{\ell^2}}{1 - ({{\cal K}/{\ell^2}})\; 
\eta_{\rho\sigma}\, x^{\rho}\, x^{\sigma}}\r)\,
\l(\eta_{\mu\nu}\, x^{\mu}\, \d x^{\nu}\r)^2,
\eeq
where the quantity ${\cal K}$ can take values $-1$, $0$ or $1$.
The symmetries of these spacetimes---viz. homogeneity and isotropy about every 
point---are more evident in, for example, the spherical coordinates, and we 
shall employ these coordinates for our calculations. 

%%%%%%%%%%%%%%%%%%%%%%%%%%%%%%%%%%%%%%%%%%%%%%%%%%%%%%%%%%%%%%%%%%%%%%%%%%%%%%%%

\subsection{The Einstein static universe}

In $(3+1)$ spacetime dimensions, the Einstein static universe (ESU), 
corresponding to ${\cal K}=1$, can be described by the line element
\beq 
\d s^{2}=-\d t^{2} + \ell^2\; \l(\d\chi^{2}+{\rm sin}^{2}\chi\, 
\d \Omega_{2}^{2}\r), 
\eeq 
where $d\Omega_{2}^{2}$ denotes the metric on a $2$-dimensional unit 
sphere. 
ESU is topologically $(\mathbb{R} \times \mathbb{S}^3)$.
The quantity $\ell$ in the above metric denotes the radius of 
$\mathbb{S}^3$, and it is related to the scalar curvature~$R$ 
of the spacetime by the relation: $R=(6/\ell^2)$.
The kernel in the ESU can be factorized as 
\beq
K_{(\mathbb{R} \times \mathbb{S}^3)}(t,{\bf x},t',{\bf x'};s) 
= \l(-4\, \pi\, i\, s\r)^{-1/2}\; 
\exp{\,-\,i\;[\l(t-t'\r)^2/4\, s]}\;\; 
K_{\mathbb{S}^3}({\bf x}, {\bf x'}; s),\label{eq:kesuF} 
\eeq
where $K_{\mathbb{S}^3}({\bf x}, {\bf x'}; s)$ is the kernel corresponding
to the coordinates ${\bf x}$ and ${\bf x'}$ on $\mathbb{S}^3$. 
Now, for $\mathbb{S}^3$, it can be shown by direct computation that 
the relation (see, for example, Ref.~\cite{camporesi-1990})
\beq
\square\, \Delta^{1/2}=\l(R/6\r)\, \Delta^{1/2}\label{eq:vvEigen} 
\eeq
holds. 
The coefficients $a_{n}$'s can then be obtained from the recursion 
relation~(\ref{eq:rr}), and they turn out to be (with $a_0=1$): 
$a_{n}= \l(1/n!\r)\, \l(R/6\r)^{n}$. 
As a result, the sum in Eq.~(\ref{eq:sde}) can be evaluated {\it exactly}\/ 
to arrive at the following form for the kernel:
\beq
K_{\mathbb{S}^3}({\bf x}, {\bf x'}; s)
= (4\,\pi\, i\, s)^{-3/2}\;\; \Delta^{1/2}\l({\bf x}, {\bf x'}\r)\;\; 
\exp\; i\,\l[\l(\sigma^{2}({\bf x}, {\bf x'})/4\,s\r) +\l(R\, s/6\r)\r],
\label{eq:kscr} 
\eeq
where $\sigma\l({\bf x}, {\bf x'}\r)$ is the geodesic distance on
$\mathbb{S}^3$.
Also, for $\mathbb{S}^3$, it can be easily shown that
\beq
\Delta^{1/2}\l({\bf x}, {\bf x'}\r) 
= \l(\frac{\sigma\l({\bf x}, {\bf x'}\r)/\ell}{\sin\, [\sigma\l({\bf x}, 
{\bf x'}\r)/\ell]}\r).
\eeq
Therefore, upon using Eqs.~(\ref{eq:kscr}) and (\ref{eq:kesuF}) 
in (\ref{eq:mGfn}), we obtain the PID modified Greens function
in the ESU to be
\beq 
G_{_{\rm PID}}\l(x, x'\r) 
= -\l(\frac{\sqrt{b}}{8\, \pi}\r)\; 
\sum^{\infty}_{n=-\infty} 
\l(\frac{H_1^{(2)}\l(\sqrt{b\, \l[u_n^2\l(x, x'\r)
- \lp^2\r]}\,\r)}{\sqrt{u_n^2\l(x, x'\r) -\lp^2}}\r)\;
\Delta_n^{1/2}\l({\bf x}, {\bf x'}\r), 
\label{eq:univ-form}
\eeq 
where $b=\l[m^2 + \l(6\, \xi - 1\r)\, (R/6)\r]$, and $H_1^{(2)}$ is the Hankel 
function of the second kind and order one.
Also, $u_n^2\l(x, x'\r)=\l[(t-t')^{2}-\sigma_{n}^{2}({\bf x}, {\bf x'})\r]$, 
and the quantities with subscript $n$ are obtained by replacing 
$\sigma({\bf x}, {\bf x'})$ with $\sigma_{n}({\bf x}, {\bf x'})
=[\sigma({\bf x}, {\bf x'}) + (2\, \pi\, n)\, \ell]$. 
Since $\mathbb{S}^3$ is compact, the sum over~$n$ in the above expression
essentially takes into account geodesics with higher and higher windings
that connect the two points ${\bf x}$ and ${\bf x'}$, while the $n=0$ term 
corresponds to the contribution due to the direct path.
In the standard case (i.e. when $\lp=0$), it is the $n=0$ term that leads 
to divergences in the coincidence limit\footnote{A quick way to see that 
the $n \neq 0$ terms are finite in the coincidence limit is to note that, 
after separating out the $n=0$ term and a series independent of $\sigma$, 
we are left with a term proportional to
\[\csc{(\sigma/\ell)} \times \sum_{n \neq 0}\, n\, 
H_{1}^{(2)}\l[i\, \sqrt{b\, \l(\sigma_n^2 + \lp^2\r)}\r]\;\; 
\l(\sigma_n^2 + \lp^2\r)^{-1/2}\]
in the modified Green's function. 
For $\sigma = 0$, the sum multiplying $\csc{(\sigma/\ell)}$ is strictly zero 
since it is odd in~$n$.
Considered as a smooth function of $\sigma$, this sum must therefore have 
an expansion in positive powers of $\sigma$, with the leading term going 
as $\sigma$. 
Therefore, in the $\sigma \rightarrow 0$ limit, we are left with a 
{\it finite}\/ contribution from this piece.}.

Therefore, we concentrate on this term to demonstrate that the PID 
modified propagator is finite in the coincidence limit. 
Using the fact that $\Delta({\bf x},{\bf x})=1$, the $n=0$ term, 
in the coincidence limit ${x}\to {x'}$, gives\footnote{The resemblance 
to the Minkowski space propagator for a massive field can be understood 
as follows.
In the flat space limit, i.e. as $\ell \rightarrow \infty$, $\Delta^{1/2} 
= 1$, as expected. 
Further, when $R$ is constant, the combination $(m^2 + \xi\, R)$ can be 
looked upon as a effective `mass' of the field.}
\beq
G_{_{\rm PID}}\l(x, x\r) 
 = -\l(\frac{\sqrt{b}}{8\, \pi\, i\, \lp}\r)\,
H_1^{(2)} \l(i\, \sqrt{b}\, \lp\r).\label{eq:coinc-limit}
\eeq
We see that the divergent term has been rendered finite by $\lp$. 
As $b \rightarrow 0$ (i.e. for the case of a massless and conformally 
coupled field), the modified propagator reduces to
\beq
G_{_{\rm PID}}\l(x, x\r) 
= \l(\frac{i}{4\, \pi^2} \r) \frac{1}{\lp^2}.
\eeq
It is clear from the {\it form}\/ of the modification that $\lp$ acts 
as a residual, {\it zero point length}\/ of spacetime intervals. 
In the same limit, for $x \neq x'$, the two point function can be
expressed as
\beq
G_{_{\rm PID}}\l(x, x\r)  
= \l(\frac{i}{4\, \pi^2} \r)
\l(\frac{\Delta^{1/2}}{u_0^2}\r)\, \l( 1 + \frac{\lp^2}{u_0^2} \r)^{-1}
\approx \l( \frac{i}{4 \pi^2} \r)\, \Delta^{1/2}\, 
\l[\frac{1}{u_0^2} - \frac{\lp^2}{u_0^4} + \cdots \r].
\label{eq:nonanalytic}
\eeq
Each term in this expression diverges as $u \rightarrow 0$, and, as a 
result, the corrections are {\it non-perturbative}\/ in $\lp$. 
Moreover, under analytic continuation, $(\sigma/\ell) \rightarrow [i\, 
(\sigma/\ell)]$, and $\ell \rightarrow (i\, \ell)$, we can obtain the 
kernel and, hence, the propagator for $(\mathbb{R} \times \mathbb{H}^3)$ 
(there is, of course, no sum over $n$ now, since $\mathbb{H}^3$ is 
non-compact, and there is a unique geodesic connecting the two points). Clearly, all the above conclusions regarding the effects of $\lp$ will
apply to this case as well.

%%%%%%%%%%%%%%%%%%%%%%%%%%%%%%%%%%%%%%%%%%%%%%%%%%%%%%%%%%%%%%%%%%%%%%%%%%%%%%%%

\subsection{de Sitter and anti de Sitter spacetimes}

The Euclidean continuation of the de Sitter spacetime is 
topologically~$\mathbb{S}^4$, and the kernel can be obtained exactly in 
such a case (see, for instance, Ref.~\cite{dowker-1976}). 
But, we find that the corresponding duality modified propagator can not 
be evaluated in a closed form using these expressions. 
So, we resort instead to the {\it Gaussian approximation}\/ to the 
kernel~\cite{bekenstein-1981}. 
Under this approximation, one constructs a Fermi normal coordinate system 
based on the extremal path, and considers only those paths in the path 
integral sum which deviate from the extremal path quadratically (in terms 
of appropriate Fourier coefficients). 
Therefore, as such, the approximation itself does not require the separations 
between the two spacetime points $x$ and $x'$ to be small. 
However, as we shall show in App.~\ref{app:vvdsn}, for $\mathbb{S}^{\mathrm N}$ 
the approximation becomes better for small geodesic separation between the 
points (which is the limit in which we are, eventually, interested).

In spacetimes of constant scalar curvature, under the Gaussian approximation, 
in $(3+1)$-dimensions, the kernel $K(x, x';s)$ can be 
expressed as ~\cite{bekenstein-1981}
\beq
K(x, x';s) 
\simeq \l(\frac{i}{(4\, \pi\, i\, s)^2}\r)\, 
\l({\rm Det}\; D \r)^{-1/2}\, 
\exp\; i\,\l[\l(\sigma^{2}(x, x')/4\,s\r) 
-i\, \l(\xi\, R\, s/6\r)\r].
\eeq
The quantity ${\rm Det}\; D$ is defined in \cite{bekenstein-1981}, where it 
has been argued that, ${\rm Det}\; D=\Delta^{-1}$ (this can be shown rigorously 
for ESU and de Sitter). 
For de Sitter, using the general result derived for arbitrary 
$\mathbb{S}^{\mathrm N}$ in the appendix, we see that,
\beq
\square \Delta^{1/2} 
\simeq \l(\frac{R}{6}\r)\, \Delta^{1/2} - \l(\frac{1}{20}\r)\,
\l(\frac{\sigma R}{12}\r)^2\, \Delta^{1/2}
- \l[{\mathcal{O}}(\sigma^4\, \ell^{-6}) + {\mathrm {higher~order~terms}} \r].
\eeq 
Hence, in the coincidence limit, $\square \Delta^{1/2} \sim (R/6) \Delta^{1/2}$ 
is indeed a good approximation, which validates the use of the Gaussian 
approximation. 
To calculate the coincidence limit, we simply note that $\Delta(x,x)=1$. 
Thus, {\it the coincidence limit of the propagator is the same as that 
for ESU},\/ given by Eq.~(\ref{eq:coinc-limit}), and is therefore finite. 
Under analytic continuation, it is easy to see that the same result would 
hold for anti de Sitter $(\mathbb{H}^4)$ as well. 
In fact, for arbitrary $\mathbb{S}^{\rm N}$, even though the exact form of 
the propagator would depend on $N$, we expect the coincidence limit to be 
finite. 
(If $\square \Delta^{1/2} \sim (R/6) \Delta^{1/2}$ is not constant, it will 
act as a `potential' and the extremal path will be \textit{accelerated}.\/ 
In this case, $\l({\rm Det}\, D \r)^{-1/2}$ becomes a complicated function 
involving $s$ \cite{bekenstein-1981}, and (unfortunately!) the propagator 
can not be obtained in a closed form in general.)

As a final comment, let us mention that in Refs.~\cite{inflation}, the
modification $\sigma^2 \rightarrow \l(\sigma^2 + \lp^2\r)$ was used in the
context of cosmology (the motivation there being the results obtained earlier 
from quantization of the conformal factor of conformally flat metrics, and not 
path integral duality). 
It was found that with this modification, one can generate density perturbations 
of acceptable magnitude, and thereby make inflation work without any fine tuning 
of parameters. Our result for the de Sitter case may be of some use in such a context.

%%%%%%%%%%%%%%%%%%%%%%%%%%%%%%%%%%%%%%%%%%%%%%%%%%%%%%%%%%%%%%%%%%%%%%%%%%%%%%%

\subsection{Anti de Sitter spacetime in $(2+1)$-dimensions}

We shall now discuss the case of the $(2+1)$-dimensional anti de Sitter
spacetime (AdS$_3$) separately, since it allows the {\it exact}\/ evaluation 
of the kernel and the modified propagator. 

Recently, we had computed the PID modifications to the propagator and 
the associated stress-energy tensor around the $(2+1)$-dimensional, 
Banados-Teitelboim-Zanelli (BTZ) black hole~\cite{dawood-2008}.
The BTZ black hole solution is obtained by discrete identifications of points
on AdS$_3$.
Hence, we had required the kernel in AdS$_3$ to arrive at the corresponding 
kernel around the BTZ black hole.
We shall be brief here, since the details can be found in our earlier work.

AdS$_3$ can be described by the line-element 
\beq 
\d s^2  = - \l(\frac{r^2}{\ell^2}-1\r)\, 
\d t^2 + \l(\frac{r^2}{\ell^2}-1\r)^{-1}\,
\d r^2 + r^2\, \d \phi^2,
\eeq
where  $-\infty < (t, \phi) <\infty$ and $0< r < \infty$. 
The kernel in such a background can be easily evaluated using the method of 
spectral decomposition, and it is found to be 
\beq
K(x,x';s) 
= \l(\frac{1}{(4 \pi\,i\, s)^{3/2}}\r)\;
\l(\frac{\sigma(x,x')/\ell}{\sinh\,
\l[\sigma(x,x')/\ell\r]}\r)\; 
\exp\, i\,\l[(\sigma^{2}(x,x')/4\, s)
-(\beta\, s /\ell^2)\r],
\eeq
where $\beta = \l[1 + (m\,\ell)^2+ \xi\, R\,\ell^{2}\r]$, $R=-(6/\ell^2)$, 
and the quantity $\sigma(x, x')$ is given by
\beq
{\rm sinh}\l[\sigma(x,x')/2\, \ell\r]
\!=\! \l(\sqrt{2}\, \ell\r)^{-1}
\biggl[-\, \sqrt{\l(r^2-\ell^2\r)\,\l(r'^2-\ell^2\r)}\;\; 
{\rm cosh}\l[\l(t-t'\r)/\ell\r]
-\;\ell^{2}+r\; r'\; {\rm cosh}\l(\phi-\phi'\r)\biggr]^{1/2}.
\eeq
From the above kernel, the duality modified Green's function in AdS$_3$ can 
be immediately obtained to be
\beq
G_{_{\rm PID}}(x,x')
= \l(\frac{1}{4\, \pi}\r)\;
\l(\frac{\sigma(x,x')/\ell}{\sinh\, 
\l[\sigma(x, x')/\ell\r]}\r)\;
\l(\frac{1}{\sqrt{\sigma^2(x,x') + \lp^2}}\r)\;
\exp{-\l(\sqrt{(\beta/\ell^2)\; [\sigma^2(x,x')
+ \lp^2]}\,\r)}.\label{eq:mGfnadS}
\eeq
In the coincidence limit, %(i.e. when ${x} \to x'$), 
$\sigma(x,x') \rightarrow 0$, and this modified 
propagator reduces to 
\beq  
G_{_{\rm PID}}(x,x) 
=\l(\frac{1}{4\,\pi\,\lp}\r)\; \exp\, -\l(\sqrt{\beta}\; \lp/\ell\r), 
\label{eq:coinc-limit-3d} 
\eeq 
which is, clearly, ultra violet finite.

%%%%%%%%%%%%%%%%%%%%%%%%%%%%%%%%%%%%%%%%%%%%%%%%%%%%%%%%%%%%%%%%%%%%%%%%%%%%%%%%

\section{Discussion}

We have analyzed the scalar field propagators in spacetimes of constant 
curvature, taking into account the Planck scale corrections according to 
the hypothesis of path integral duality. 
The main results can be summarized as follows:
\begin{enumerate}
\item
In $(3+1)$-dimensions [cf. Eqs.~(\ref{eq:univ-form}) 
and~(\ref{eq:coinc-limit})]:
\bea
G_{_{\rm PID}}(x, x') 
&=& -\l(\frac{\sqrt{b}}{8 \pi}\r)\; 
\l(\frac{H_{1}^{(2)}\l(\sqrt{b\, 
\l[u^2(x,x') -\lp^2\r]}\,\r)}{\sqrt{u^2(x,x')
- \lp^2}}\r)\; \Delta^{1/2}(x,x')\nn\\ 
&\underset{x \rightarrow x'}{\longrightarrow}& 
-\l(\frac{\sqrt{b}}{8\, \pi\, i\, \lp}\r)\, 
H_1^{(2)} \l(i\, \sqrt{b}\, \lp\r),\nn
\eea
\item
In $(2+1)$-dimensions [cf. Eqs.~\ref{eq:mGfnadS}) 
and~(\ref{eq:coinc-limit-3d})]
\bea
G_{_{\rm PID}}(x,x') 
&=& \l(\frac{1}{4\, \pi\, \ell}\r)\; \Delta^{1/2}(x,x')\; 
\l(\frac{\exp{-\sqrt{b\, \l[u^2(x,x')-\lp^2\r]}}}{\sqrt{u^2(x,x')
- \lp^2}}\r)\nn\\ 
&\underset{x \rightarrow x'}{\longrightarrow}& 
\l(\frac{1}{4\, \pi\, \lp}\r)\, \exp{-\l(\sqrt{b}\; \lp\r)},\nn
\eea
\end{enumerate}
where $u(x,x')$ represents the geodesic distance between the two points $x$ and
$x'$, and the other symbols have their meanings as stated in the text (multiple 
geodesics must be accounted for, as the case may be). 
We have therefore shown that, the hypothesis of PID, when applied to cases where 
the proper time kernels are known exactly (or under a suitable approximations): 
(i)~regulates the theory at Planck scales, (ii)~yields modifications which are
non-perturbative in~$\lp$ and, (iii)~most interestingly, we find that the quantum 
gravitational effects, as accounted for by the duality prescription, can be looked 
upon as leading to addition of $\lp$ to {\it all}\/ spacetime (geodesic) intervals 
in a (peculiar) Pythagorean way, that is, $\l\langle \sigma^2(x,x')\r\rangle 
= [\sigma^2(x,x') + {\cal O}(1)\, \lp^2]$, as is clearly evident from the above 
expressions for the propagators.

The duality invariance of the relativistic point particle path integral 
is therefore {\it equivalent}\/ to `adding' a zero-point length to spacetime 
intervals, which is a non trivial point (as was already emphasized in 
\cite{paddy-1997-1998}). 
Such a result might be an outcome of some {\it generic short distance 
behavior}\/ of the spacetime structure itself, when quantum effects are 
taken into account. 
One expects that these results would naturally appear in the (effective) 
low energy sector of the full theory of quantum gravity. 

%%%%%%%%%%%%%%%%%%%%%%%%%%%%%%%%%%%%%%%%%%%%%%%%%%%%%%%%%%%%%%%%%%%%%%%%%%%%%%%

\section*{Acknowledgments}

DK is supported by the Senior Research Fellowship of the Council for Scientific and
Industrial Research, India. LS would like to thank the Inter University Centre for
Astronomy and Astrophysics, Pune, India for hospitality, where part of this work was 
carried out. SS is supported by the Marie Curie Incoming International Grant IIF-2006-039205.

%%%%%%%%%%%%%%%%%%%%%%%%%%%%%%%%%%%%%%%%%%%%%%%%%%%%%%%%%%%%%%%%%%%%%%%%%%%%%%%
%%%%%%%%%%%%%%%%%%%%%%%%%%%%%%%%%%%%%%%%%%%%%%%%%%%%%%%%%%%%%%%%%%%%%%%%%%%%%%%
\appendix
%%%%%%%%%%%%%%%%%%%%%%%%%%%%%%%%%%%%%%%%%%%%%%%%%%%%%%%%%%%%%%%%%%%%%%%%%%%%%%%
\section{Appendix: Conformal factor as a physical degree 
of freedom}\label{app:3d-conformal}

One often finds statements in literature that the conformal factor enters in 
the action with the `wrong' sign in the kinetic term, and therefore must be 
unphysical. 
This is incorrect, as is easily seen from the following arguments.

Consider the general metric (for the sake of convenience, we shall work below 
in $(3+1)$-dimensions)
\beq
\d s^2 = - N^2\, \d t^2 + 2\, N_{j}\, \d t\, \d x^j 
+ h_{i j}\, \d x^i\, \d x^j,\label{eq:metric-gauge}
\eeq
with the gauge choice, $N=\Omega$, $N_{j} = 0$, and $h_{i j} = (\Omega^2\,
\gamma_{i j})$ (the metric components can, in general, depend on all the 
coordinates). 
In such a case, the metric reduces to
\beq
\d s^2 = \Omega^2\, \l(- \d t^2 + \gamma_{ij}\, \d x^i\, \d x^j\r)
= \Omega^2\; g_{\mu \nu} \; \d x^{\mu}\, \d x^{\nu}.
\eeq
Using the relation between the Ricci scalars of conformally related metrics, 
it is easy to show that the Einstein-Hilbert action can be written as
\beq
S_{_{\rm EH}} = \l(\frac{1}{16\,\pi\,G}\r)\,
\int \d t\; \d^3 x\; \sqrt{\gamma}\; \l[-6\, \Omega\, \Box_g\, \Omega + R\r],
\eeq
where $\Box_g \Omega =\l(- \partial_t^2\, \Omega + \gamma^{i j}\, \partial_i\, 
\partial_j\, \Omega\r)$, and $R$ is the Ricci scalar of the metric $g_{\mu \nu}$. 
The action can be rewritten as
\bea
S_{_{\rm EH}} 
&=& \l(\frac{1}{16\,\pi\,G}\r)\,
\int \d t \; \d^3 x \; \sqrt{\gamma}\; 
\l[6\, \Omega\, \partial_t^2\, \Omega - 6\, \Omega\, 
\gamma^{i j}\, \partial_i\, \partial_j\, \Omega + R\r]\nn\\
&=& \l(\frac{1}{16\,\pi\,G}\r)\;
\int \d t \; \d^3 x \; \sqrt{\gamma} \; 
\l[-6\, \l( \partial_t \Omega \r)^2 + \ldots \r],
\eea
where we have exhibited only the kinetic term associated with $\Omega$. 
We immediately see that the kinetic term has the wrong sign, which is the 
reason why the conformal degree of freedom is often regarded as unphysical. 

That this is incorrect can be easily seen by choosing instead the gauge, 
$N=1$, $N_{j} = 0$ and $h_{ij} = \l[F^{2}(x)\, \gamma_{i j}\r]$ in 
Eq.~(\ref{eq:metric-gauge}), so that the metric becomes
\beq
\d s^2 = - \d t^2 + F^2\, \gamma_{i j}\, \d x^i \,\d x^j.
\eeq
In this case, $F$ generates a class of metrics with \textit{3-geometries}\/ 
conformal to $\gamma_{i j}$, which is chosen to have, say, ${\rm Det}\, 
\gamma= {\rm constant}$. 
Now, we know that, the components of the 3-metric definitely correspond to 
physical degrees of freedom. 
However, explicitly calculating the action, we find that
\beq
S_{_{\rm EH}} 
= \l(\frac{1}{16 \pi G}\r)\, 
\int \d t \; \d^3 x \; 
\sqrt{\gamma} \; \l[- 6\, F\, \l(\pa_t F \r)^2 + \ldots \r].
\eeq
The kinetic term corresponding to $F$ again appears with the wrong sign, 
even though we know that $F$ in this case corresponds to a physical degree 
of freedom. 
This clearly illustrates that the conformal factor is \textit{not}\/ an 
unphysical degree of freedom.

%%%%%%%%%%%%%%%%%%%%%%%%%%%%%%%%%%%%%%%%%%%%%%%%%%%%%%%%%%%%%%%%%%%%%%%%%%%%%%%
\section{Appendix: $\square \Delta^{1/2}$ for 
$\mathbb{S}^{\mathrm N}$} \label{app:vvdsn}

For a unit $N$-sphere, the Van-Vleck determinant is given 
by~\cite{camporesi-1990}
\beq
f(\sigma) \equiv \Delta(x,x')^{1/2} 
= \l(\frac{\sin{\sigma(x,
x')}}{\sigma(x,x')} \r)^{{(N-1)}/{2}}
\eeq
(For $\mathbb{H}^{\mathrm N}$, $\Delta^{1/2}=\l( \sinh{\sigma}/\sigma \r)^{(N-1)/2}$;
all the steps below go through, with a few sign changes, while the final expression 
remains the same.)
We begin by writing the $\square$ operator in Riemann normal coordinates as (see, for 
example, Ref.~\cite{parker-cargese})
\bea 
\square = \frac{\partial^2}{\partial
\sigma^2} + \l[\frac{\partial}{\partial \sigma}
\l(\ln\,{\Delta^{-1}}\r) + \frac{N-1}{\sigma} \r]\,
\frac{\partial}{\partial \sigma}
\eea
It then follows that
\beq
\square f(\sigma) = \frac{(N-1)\,(N-3)}{4}\,
\l[\frac{1}{\sigma^2} 
-\frac{1}{\sin^2{\sigma}} + \frac{N-1}{N-3} \r]\, f(\sigma)
\eeq
We now expand the terms within the square bracket in the limit
$\sigma \rightarrow 0$, and obtain
\bea
\square f(\sigma) 
&=& \frac{(N-1)\,(N-3)}{4} \, 
\l[\frac{2\, N}{3(N-3)} 
- \frac{\sigma^2}{15} - {\mathcal O}(\sigma^4)\r]\, f(\sigma)\nn\\
&=& \l[\frac{N\, (N-1)}{6} - \frac{(N-1)\,(N-3)}{60} \, \sigma^2 -
{\mathcal O}(\sigma^4)\r]\, f(\sigma)\nn\\
&=& \l(\frac{R}{6} \r)\, f(\sigma) + {\mathcal O}(\sigma^{2 k})
\hspace{0.5in} \l(k=1, 2,\ldots \r).
\eea
In the last step, we have used the fact that, for $\mathbb{S}^{\mathrm N}$ of 
unit radius, $R=[N\, (N-1)]$. 
Note that all terms of ${\mathcal O}(\sigma^2)$ and higher vanish for $N=3$, 
which is the case encountered in ESU ($\mathbb{R} \times \mathbb{S}^3$). 
The Gaussian approximation is therefore \textit{exact}\/ for 
$\mathbb{S}^3$ \cite{bekenstein-1981}. 
This result suggests generalizations of arguments used for ESU and de Sitter 
in $(3+1)$-dimensions, to arbitrary $\mathbb{S}^{\mathrm N}$ or 
$\mathbb{H}^{\mathrm N}$.

%%%%%%%%%%%%%%%%%%%%%%%%%%%%%%%%%%%%%%%%%%%%%%%%%%%%%%%%%%%%%%%%%%%%%%%%%%%%%%%

%%%%%%%%%%%%%%%%%%%%%%%%%%%%%%%%%%%%%%%%%%%%%%%%%%%%%%%%%%%%%%%%%%%%%%%%%%%%%%%
\end{document}